%% file: TopArgelander.tex
\documentclass[apj,numberedappendix]{emulateapj}
\pdfoutput=1
\usepackage{ulem}
\usepackage{epstopdf}
\usepackage{hyperref}
\usepackage{threeparttable}
\usepackage{pdfpages}
\usepackage{csquotes}
\usepackage[multiple]{footmisc}

\def\gtorder{\mathrel{\raise.3ex\hbox{$>$}\mkern-14mu
             \lower0.6ex\hbox{$\sim$}}}
\def\ltorder{\mathrel{\raise.3ex\hbox{$<$}\mkern-14mu
             \lower0.6ex\hbox{$\sim$}}}

\slugcomment{Draft of \today}
\shorttitle{Top Argelander Stars}
\shortauthors{S.\ R.\ Kulkarni}

\begin{document}
\title{Top Argelander Stars: Pedagogy \& Prize}
\author{S.\ R.\ Kulkarni}

\begin{abstract}
 Stellar astronomy, fueled by massive capital investments, advances
 in numerical modeling and theory, is resurgent and arguably is on
 the verge of a magnificent renaissance. Powerful time domain optical
 surveys, both on ground and in space, are producing data on variable
 stars on an unprecedented  industrial scale. Those with deep
 knowledge of variable stars will stand to benefit from this
 resurgence.  Notwithstanding these developments, in some astronomical
 communities,  classical stellar astronomy has been in the doldrums.
 I offer a modest proposal to establish a basic level of familiarity
 with variable stellar phenomenology and an attractive scheme to
 make research in variable star astronomy visible, alluring and
 fashionable.
\end{abstract}

\section{A Resurgence in Stellar Astronomy}

Stellar astronomy is now on the rise. {\it Kepler} with its exquisite
photometric precision has revolutionized the field of extra-solar
planets and additionally has kick-started the rich field of
astero-seismology.  The {\it Gaia} mission is poised to  initiate a
revolution in astrometry on a scale that astronomers have not seen
for nearly a century.  This grand expectation is  grounded in the
large number of stars for which precision astrometric measurements
will be obtained by {\it Gaia}.  Pan-STARSS has laid down a deep
multi-band photometric grid over three quarters of the sky.  Next
year, the Zwicky Transient Facility (ZTF) will begin a 3-year survey
of the Northern sky, sharply focused  on the dynamic optical sky
(that is, variable stars, transients and moving objects). In 2018
the TESS mission is expected to start routine observations. TESS
will focus on precision photometry of stars brighter than 12 mag.
However, full frame images will allow observations of fainter stars
over substantial areas of the  sky.

One should not discount smaller and nimble investments such as ASAS,
ASAS-SN and Evryscope. Many such projects are trail blazers. Finally,
it is both gratifying and astonishing, despite
the massive capital investment by professional astronomers, 
 that amateur astronomers continue to make serious contributions to stellar and supernova
research.

While above we have focused on astrometry and photometry there has
been good progress in spectroscopic surveys (e.g.\ RAVE and the
ultra-low resolution spectroscopy of {\it Gaia}). There is now a
(timely) movement to repurpose the massively-multiplexed spectrographs
of the Sloan Digital Sky Survey (SDSS) for stellar astronomy.

As can be deduced from the above summary the capital investment in
stellar astronomy has been substantial.  However, stellar astronomy
is a fairly mature field. Thus merely obtaining large volumes of
superior data without corresponding superior theoretical framework
will  not result in realization of the full gains of the capital
investments.  Fortunately, there has been good progress on the
theoretical and modeling fronts.  Nowadays, on a routine basis,
astronomers undertake multi-dimensional, if not three-dimensional,
modeling of stars and related phenomena. Tools such as MESA  allow
observers and modelers to compare theory to observations at the few
percent (or even lower) level.

The future landscape in stellar astronomy is extremely promising.
I justify this statement with just three examples.  First is the
implication to stellar astronomy arising from the recent findings
by LIGO \citep{2016ApJ...818L..22A} -- not only are stellar black
hole binaries abundant but the masses of the black holes was
surprisingly higher than expected.  These two findings are motivating
astronomers to explore entirely {\it new paths of stellar evolution}
in which not only low metal abundance but rapid rotation is expected
to strongly influence the evolution of stars.  Second,  several
findings over the last few years,  from time domain surveys -- wild
diversity in pre-supernova mass loss, new types of super-luminous
supernova -- clearly show that massive stars have multifarious
evolutionary paths and multifaceted deaths.  These advances are
still being digested. It is now clear that future textbooks on
stellar astronomy will have not one but several chapters on the
evolution and endings of massive stars. Finally, it is becoming
clear that massive stars in binary systems may play a major role
in the ionization history of the early Universe ($z\sim 6$;
\citealt{mkh+15}).

\section{F.\ W.\ Argelander}

With the modern background laid out I now arrive at the primary
goal of this writeup -- the anticipated resurgence of variable star
astronomy.  It is by now well known that forecasting the future is
difficult\footnote{\url{http://www.famous-quotes-and-quotations.com/yogi-berra-quotes.html}}. Furthermore, it
is almost always wrong.\footnote{\url{http://www.smithsonianmag.com/smart-news/why-experts-are-almost-always-wrong-9997024/?no-ist}}  However, it is said that a knowledge of
the past helps understand the future better. In that spirit our
journey begins with Friedrich Wilhelm Argelander (1799--1875) who
is considered as the founding astronomer of the modern era of
variable stars.

Argelander was born in what is now Lithuania. He undertook his PhD under
the famous German astronomer Friedrich Bessel at the University of
K\"onigsberg located in what is now called Kalingrad.\footnote{The modern
name for this University is Immanuel Kant Baltic Federal University.} Following his PhD,
in 1823, 
he moved to the Turku Observatory.  In 1832 he moved to Helsinki where  he supervised the construction of the then
new Helsinki observatory. His primary
accomplishment, up until this point, was the investigation of the Sun's motion with
respect to other stars. In 1837 he moved to Germany and took up the 
Directorship of the Observatory of the University of Bonn (which was, at the 
time, one of the best funded observatories).  It was in Bonn, in 1844, he began studies
of variable stars. Separately,
since 1826, he was a corresponding member of the St.\ Petersburg Academy of Sciences.
Clearly, Argelander was not only energetic but also a pan-European astronomer.

Argelander was an expert in astrometry and photometry. Along with
his students and colleagues he was responsible for the famous star
catalog, {\it Bonner Durchmusterung}\footnote{German for ``thorough
check''} (BD).  This catalog, consisting of 324,188 stars with
magnitude brighter than about nine, had an  astrometric precision
was $0.1^\prime$ in declination and 0.1\,second in Right Ascension (RA). It was the first comprehensive astrometric
and photometric catalog  of the Northern sky. The catalog,
for its time, was considered to have superior photometry.

The BD catalog laid the standard for future photometric and astrometric
star catalogs.\footnote{The BD catalog, given the $51^\circ$ latitude
of Bonn, was deficient in Southern stars, $\delta<-2^\circ$. This motivated the {\it
Cordoba Durchmusterung} (CD) and the {\it Cape Photographic
Durchmusterung} (CPD) catalogs.}  The BD catalog was re-issued in 1950! Many BD
stars serve as absolute photometric standards for ground- and
space-based telescopes (e.g.\ BD\,+28$^\circ$\,4211,
BD\,+17$^\circ$\,4708). 
Separately, 
Argelander founded the Astronomische Gesellschaft (Astronomical Society), which in collaboration with many observatories expanded his work to produce the AG catalogs.\footnote{\url{https://www.britannica.com/biography/Friedrich-Wilhelm-August-Argelander}}

\subsection{Variable Stars \&\ Argelander}
\label{sec:Argelander}

A small  note in {\it Popular Astronomy} by Annie J. Cannon of the
Harvard College Observatory \citep{1912PA.....20...91A}  written
in 1911 is an excellent gateway to appreciate the role played by
Argelander in the development of variable star astronomy.  In that
note Cannon refers to a chapter on variable stars, entitled {\it
An appeal to the friends of Astronomy}, written by Argelander  for
the 1844 edition of Schumacher's Astronomical Year book.  Argelander
exhorts fellow astronomers to undertake observations of variable
stars and catalog the photometry.  Cannon notes ``After reading the
article, one feels like reviewing the advances in our knowledge of
the variable stars since 1844. Instead of 18 variables, as in
Argelander's catalogue, we must provide for more than 4000."

Argelander was interested in more than simply amassing data and
identifying variable stars.  Even with the meager sample of variable
stars at his disposal, Argelander in the afore mentioned chapter
speculated about the physical origins of variability, making the
following prescient suggestions:

\begin{displayquote} \textit{On account of the low state of our
knowledge of these stars, nothing in general can at present be
offered nor, by any means, can a definite theory be given, which
can refer the light changes to any one cause. But happily, hypotheses,
even if full of error, fail us not. Omitting those which at first
glance are seen to be untenable, they resolve themselves into the
following three.}

\begin{enumerate}
    \item \textit{Revolution of the stars on their axes, their
    surfaces being of different luminosity on the different sides,
    whereby they would be brighter if they turned towards us the
    side of greatest illumination, or conversely, darker if the
    side of less illumination.} \item \textit{Revolution on their
    axes, with strongly compressed figure, and considerable variation
    of angle of the axis of rotation towards the line of sight. If
    the axis nearly coincides with the line of sight, then the star
    turns towards us a very extensive surface, sends us much more
    light, and therefore shines brighter than if they, because of
    a very large angle, turn their edge, if I may so call it.} \item
    \textit{Huge planets revolving around the stars, in the plane
    of whose orbits the line of sight nearly falls and which,
    therefore, by inferior conjunction with the star, cut off a
    large part of the light formerly coming from it to us, so that
    it seems less bright.}
\end{enumerate}

\textit{The first of these hypotheses seems to be the most plausible
and, in general, to explain observed appearances of several of the
stars, if we assume that the constitution of these stars is similar
to that of our Sun.} \end{displayquote}

Argelander's suggested phenomena can be summarized as: (1) rotational
modulation of star spots\footnote{Here Argelander points to the
observations of sunspots from Herschel, Schr{\"o}ter, and S{\"o}mmering,
though at that time mountain chains were considered a plausible
explanation for the spots.}, (2) precession of a rapidly rotating
ellipsoidal star, and (3) transits of extrasolar planets. Although
none of Argelander's eighteen original variables fit his hypotheses\footnote{Most
are Mira-type pulsators, Cepheids, or semi-detached eclipsing
binaries, which do exhibit ellipsoidal modulation though not due
to precession of a single star.}, star spots and transiting extrasolar
planets are today commonly observed phenomena. Doubtless, Argelander
stood on the shoulders of giants; observations of transits within
our solar system date back to at least 1631, when Pierre Gassendi
recorded the transit of Mercury, while Immanuel Kant and Pierre-Simon
Laplace are often credited for suggesting that planets should also
exist around other stars just a generation before
Argelander\footnote{Though this idea may even go back to Democritus.}.
Nevertheless, Argelander's suggestion that transits of extrasolar
planets might be observed is apparently the first such mention in
the literature.  In contrast to his prescience in astronomy,
Argelander appears to have been an unsophisticated  (simple?) person
in real life.\footnote{My colleague Prof.\ Franciscus Wilhelmus
Maria Verbunt, University of Nijmegen, The Netherlands, informed
me: ``There is a nice anecdote about Argelander. I read this in the
book on Greek Cultural History by Burckhardt, in the introduction,
where it is told by Kaiser Wilhelm himself(!): Kaiser Wilhelm was
in Bonn and decided to visit the observatory. `And, my dear
Argelander', he said, ``What is new in the starry skies?". To which
Argelander answered: ``Does your majesty already know the old?" "}

I end by noting that this section, for most part, was contributed
by graduate student Trevor David and Prof.\ Lynne Hillenbrand, both
of my own department.

\section{A Scheme to Name Variable Stars}

Thanks to Argelander's enthusiasm for variable star astronomy
research and perhaps more importantly to the precision of the
photometry in the BD catalog Argelander realized that the sky could
be  teeming with variable stars. The old scheme of assigning
interesting stars (not merely variable) with romanized Latin alphabets
had a capacity for 23 stars.\footnote{The
romanized Latin alphabets did not include J, U and W. The alphabet
I and J are variants and also not pronounced uniformly in European
languages. An astute person may have noticed that the ``I'' column
is excluded in airplane seat names.  The reader may wish to ascertain
this claim the next they fly on an airplane.}   In order to accommodate larger
populations of variable stars Argelander designed a new naming
scheme.  The scheme was reformulated several times as the number
of cataloged variable stars increased.  The reader is referred to
\citet{1915PASP...27..209T}
for a historical account of the evolution
of the naming framework.

The naming formulation has three steps and is explained below.
 \begin{itemize}

\item[I\ ] The designation of a variable star is a prefix to the
name of the constellation in which the variable star is located.
The first prefix is R.\footnote{Since the previous alphabets of
Romanized Latin were already assigned to special stars, e.g.\ the
will known P\,Cygni and Q\,Cygni (Nova\,Cygni\,1867).} As additional
stars were discovered they were assigned prefixes S, T, U, V, W,
X, Y and Z. These 9 entries, per constellation, constitute the first
(``I") series.  Examples include ``R\,Coronae Borealis" (exotic
carbon rich star), ``S\,Andromedae'' (supernova of 1885 in the
Andromeda galaxy) and ``T\,Tauri'' (the archetypical pre-main
sequence star).

\item[II\ ] However, by 1881, the number of variable stars in some
constellations already exceeded nine. In 1881 Hartwig proposed
adding the next series:  RR to RZ, SS to SZ, TT to TZ and finally
ZZ.  Thus  series II has 9+8+7+6+5+4+3+2+1 stars or a total of 45
stars.  For instance, ``RR\,Lyrae'' is an exemplar of a class of
pulsating variables. ``RS\,CVn'' is a prototype of stars with strong
magnetic and related activity powered by binarity.``ZZ\,Ceti'' is
famous for being the first white dwarf seismological pulsator.

\item[III\ ] By 1904, the number of variable stars had increased
(particularly in the Orion constellation) to such an extent that a
new series was added: AA through AZ, BB to BZ and end with QQ to
QZ.  However, as noted earlier, the letter ``J" was excluded. The scheme ends with
QZ. With some care it can be shown that the number of prefixes in
series III is  $25+24+...11+10$ or 280 stars.

Adding the number of entries from series I, II and III leads to 334
prefixes. We will call such stars as ``classic" Argelander stars.

\item[IV\ ] The advent of photography vastly increased the number
of variable stars. When the classic 334 prefixes were used up the
scheme switches to the ``modern'' V (for variable) numbering scheme.
This scheme was suggested by several astronomers including Townsley
({\it ibid}).  The scheme  starts with ``V335'' and marches to
larger numbers.  For example the variable after ``QZ\,Cyg'' (an
``irregular'' variable) is ``V335 Cyg'' (an M1 variable star).
Fortunately, mathematicians inform us that there exists  an infinite
supply of integer numbers.  Thus astronomers can safely expect no
future reformulation of the scheme to name variable stars.

\end{itemize}

It is important to note that the Argelander designation was based
on optical variability. As  a result  some super-famous variable
stars of our age have lowly Argelander designation. I quote some
examples.  ``V404 Cyg'' is a famous X-ray nova and very much in the
news since its burst in 2015.  Aquila X-1 (V1333\,Aql) is a famous
soft X-ray transient\footnote{An object which, in my youth, I
intensively observed searching in vain for the disappearance of the
accretion disk and emergence of radio pulses.}. SS\,433, an exotic
and unique, to date, stellar system\footnote{This system propelled
Bruce Margon to stardom.} is merely V1343\,Aql.

\subsection{Criticism of the Argelander Scheme}
\label{sec:Criticism}

In contrast to the effusive praise of Argelander by Cannon Townsley
was critical of Argelander for his choice of naming scheme. There
is some merit in Townsley's dislike of the convoluted and idiosyncratic  naming scheme
described above. However, from personal experience I know that
naming schemes (1) rarely have a foundation in some ``rational"
framework and (2) invoke strong emotional response from otherwise
reasonable people.   Rather than distract the reader for the main
topic of this article I refer the interested reader
\S\ref{sec:NamingSchemes} for my experience and thoughts on naming
schemes.

Returning to the topic of this article it is simply the case that
we can remember short names. It is equally true that we cannot
remember long names, even if constructed on a  rational basis.
Below, I provide one specific example to illustrate this point.


Some time ago I got interested in compact double degenerates. Even
within this group of interesting sources the ROSAT source RX\,J0806.3+1527
(sometimes shortened to RX\,J0806+15) is extremely interesting. It
has an orbital period of only 5.4 minutes. I had a very hard time
remembering the name of the source (other than remembering it is
an ``8-hour" source).  Every time I had to look up the literature
on this source I would consult a specific paper (which I could remember since one of the
authors was my friend) which referred to the source and then passed the ROSAT
source name to  SIMBAD to find recent papers on the source.

The ROSAT designation, RX\,J0806.3+1527, was consistently used since
discovery \citep{br99,ipc+99} until 2007 when \citet{bmd+07} used
an Argelander designation, HM\,Cancri.
The reader will undoubtedly agree that it is much easier
to  HM\,Cnc than RX\,J0806.3+1527.  Thanks to this short name, practically
a nemonic to me,  I can
now query  the Astrophysical Data System (ADS)\footnote{\url{http://www.adsabs.harvard.edu/}} using the  SIMBAD object filer (with name set to
HM\,Cancri) and find all the papers related to this source.  The
``upgrade" of RX\,J0806.3+1527 to HM\,Cancri has increased my
productivity!  

Finally, as with many other aspects in our culture,
conventions once established, however arcane, are hard to uproot and one may as well
as celebrate tradition rather than complain.

I end this section by condensing the history of variable star
research post Cannon's era.  After the second World War the
International Astronomical Union (IAU) gave the responsibility of
maintaining the catalogue of variable stars to two groups in the
Soviet Union (Moscow University and the Academy of Sciences).  The
two teams painstakingly maintained the rapidly growing lists of
variable stars (with new entries from non-optical bands, particularly
X-ray missions). The task involved obtaining the most accurate
positions, cross-matching of names and classification (determining
to which class each variable belongs to).  The ``General Catalogue
of Variable Stars" (GCVS)\footnote{Currently centered at the Sternberg
Astronomical Institute of the Lomonosov Moscow State University of
Russia; see \url{http://www.sai.msu.su/gcvs/gcvs/}} was the primary
reference for practitioners of  stellar variable research.  GCVS
is now linked primarily to Varbial Star Index (VSX)\footnote{\url{https://www.aavso.org/vsx/}}.
However, like many professional astronomers I use 
SIMBAD\footnote{\url{http://simbad.u-strasbg.fr/simbad/}}
because it has links to catalogues (especially those published as
accompaniments of papers) and also papers (via
ADS).

\section{The Top Argelander stars: A Pedagogical Tool}

Two years ago I taught a course on High Energy Astrophysics.  I
started the white dwarf teaching module by describing the gradual
identification of a curious star (Sirus\,B) and the puzzle it posed
to astronomers, especially to the then doyen of astronomy, Arthur
Eddington.  This was then followed by the usual discussion of
degeneracy pressure, polytropic solutions and the rich and
interesting physics of white dwarf cooling.

Inspired by a popular cultural
practice\footnote{\url{https://en.wikipedia.org/wiki/Casey_Kasem}} of
``top ten in the last ten years" I  compiled a list of top ten white
dwarfs (ranked by the number of papers attributed to them).  I then
asked the students to read up on the literature\footnote{The
\texttt{wikipedia} is usually a good starting point.} and then have
each student write up a report and deliver a presentation on the
white dwarf that caught their attention or piqued their interest.
I believe that the experiment was a success in that the students
were able to proceed to the next level of education, namely develop
a sound understanding of the  phenomenology of the subject.  My
colleague and good friend E.\ Sterl Phinney implemented the same
scheme for an undergraduate class  but for neutron stars. He too
reported success.

\begin{deluxetable}{lll}
\tabletypesize{\scriptsize}
\tablecaption{Argelander Stars (and quasi stars) ranked by publications}
\tablewidth{0pt}
\tablehead{
\colhead{Star} & \colhead{Papers} & \colhead{Citations}
}
\input{TableTopStars}
\tablenotetext{}{\small Column 1: star name. Column 2: number of
papers attributed to object by SIMBAD. Column 3: number of citations
attributed to object by ADS.}
 \label{tab:TopStars}
\end{deluxetable}

In the spirit discussed above I decided that rank ordering the
Argelander stars would be of some value. After all the Argelander
stars are the brightest variable  stars in the optical sky. Thus,
any astronomer who wishes to undertake research in variable star
astronomy would clearly benefit from being familiar with the popular
(top ranked) Argelander stars.

\subsection{Data Generation \& Results}

The sky, following an IAU resolution in 1919, is divided into 88
constellations. As noted in \S\ref{sec:Argelander}, in a given
constellation, there can be up to maximum of 334 variable stars
with Argelander prefixes.  Thus the total number of classical
Argelander stars over the entire sky is $88\times 334=29,392$.

I wrote a short program in MATLAB to inquire\footnote{The exercise
was carried out on 25 September 2016.} SIMBAD the details of each
of  these 29,392 possibilities.  Using standard Unix tools I filtered
the outputs returned by SIMBAD  and extracted the number of papers
(``references" in the lingo of SIMBAD) for each classical Argelander
star. With the $88\times 334$ matrix now populated I was in  position
to undertake a number of analyses.

I immediately noticed that not all the 29,392 possibilities had
SIMBAD entries. For instance, in Antlia there are no Argelander
stars beyond CF\,Ant. We will revisit this issue in \S\ref{sec:Prize}.
Next,  no papers were listed in SIMBAD for  991 stars.  Examples
include AY\,And, BF\,And, DD\,Ara, PP\,Vul etc.  The SIMBAD websites
states clearly ``Simbad bibliographic survey began in 1950 for stars
(at least bright stars) and in 1983 for all other objects (outside
the solar system)". However, a check of AY\,And shows two references
in the GCVS (Ref\#00150=N.Florja, Perem Zvezdy 5, 258, 1940;
 Ref\#03188 = M.Doeppner, MVS N575, 1961).  Finally, I noticed an
anomaly.  In a few constellations, there are gaps in the usage of
the prefixes, e.g.  in Andromeda all prefixes except VV have been
used.  This puzzle has now been solved, thanks to efforts on the
part of some sincere colleagues (see \S\ref{sec:VVAnd}).

Once the data gathering was finished I undertook some analysis.
The first exercise was to simply sort the stars in descending order
of the number of papers.  For the top one hundred stars thus ranked
I had ADS queried for the citations to each star.  Fifty of the
stars ranked by the number of papers along with the corresponding
number of citations are listed in Table~\ref{tab:TopStars}.  Next,
I re-sorted the list of hundred stars, this time by the number of
citations.  In Table~\ref{tab:TopCites} I list the top fifty stars
rank ordered by citations and provide the corresponding number of
papers. Next, in Table~\ref{tab:TopThree}, I list the top three
stars in each constellation. Perhaps even for an academic this is
an artificial exercise since many modern astronomers do not organize
their research by constellations. However, the last column in
Table~\ref{tab:TopThree} is quite interesting and motivated the
next section (\S\ref{sec:Prize}).

\begin{deluxetable}{lll}
\tabletypesize{\scriptsize}
\tablecaption{Argelander Stars (and quasi stars) ranked by citations}
\tablewidth{0pt}
\tablehead{
\colhead{Star} & \colhead{Citations} & \colhead{Papers}
}
\input{TableTopCites}
\tablenotetext{}{\small Column 1: star name. Column 2: number of
citations attributed to object by ADS. Column 3: number of papers
attributed to object attributed by  SIMBAD.} 
 \label{tab:TopCites}
\end{deluxetable}

\subsection{Some Remarks}

At this point a reader could reasonably expect a Reader's Digest
summary for each of these stars. However, I specifically avoid doing
so because the point of this write up is in fact to motivate
(inspire?) young astronomers to pursue a wide knowledge of astronomical
phenomenology. Such breadth can only be be earned via hard work
(plain old curiosity, attending colloquia without texting, reading
papers without intermittently checking email  and occasionally
thinking).  I would like to imagine that a student who is interested
in stellar astronomy will be puzzled by the top-ranked Argelander
star, CM\,Tau (hint: it is located in Messier~1).  More seriously,
I hope that a student who has read this writeup will make up his/her
deficiency in education by reading key literature on the stars (the
number to be decided by the student) listed in the Tables.
Alternatively, a scholar could organize a one-day event -- ``Argelander
Jamboree".  On this day  earnest neophytes, after having read up
on the literature of say thirty stars, meet and hear experts make
short presentations on each star.

Despite stating, in the previous paragraph, my desire for reticence
I am compelled to make two remarks. The stars listed in
Table~\ref{tab:TopStars} encompass an astonishing range of astronomical
phenomena.  The list includes the fundamental stellar families
(nuclear burning, degenerates and collapsed stars). Next, some stars
in Table~\ref{tab:TopStars} are extra-galactic (AGN, quasar, blazar).
Finally the list spans the full life cycle of stars, from nurseries
to death (and afterlife).

\begin{figure}[htbp]
 \centering
  \includegraphics[width=2.5in]{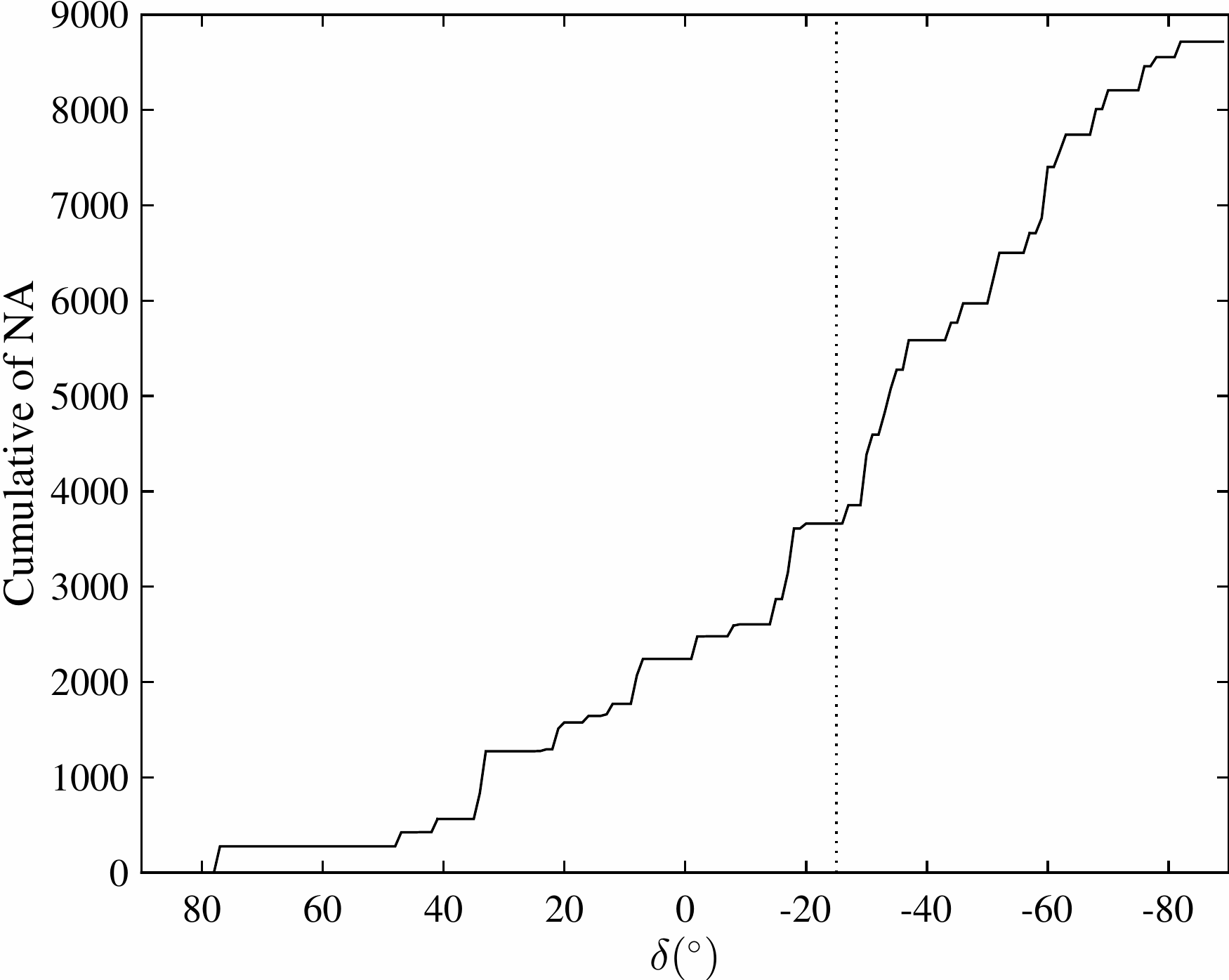}
   \caption{\small
    The cumulative number of ``not assigned" (NA) classic Argelander
    stars as a function of declination.  The vertical line
    $\delta=-25^\circ$. Data drawn from Table~\ref{tab:TopThree}.
    }
 \label{fig:CumulativeNA}
\end{figure}

\section{Argelander Designation as a Prize}
\label{sec:Prize}

In Table~\ref{tab:TopThree}  I list the number of Argelander
designations that have not been assigned (``NA").  Bearing in mind
of the improvement in productivity when RX\,J0806.3+1527 was given
an Argelander name (\S\ref{sec:Criticism}) and noting the large
vacancies in Table~\ref{tab:TopThree} an idea emerged in my mind
-- use the remaining Argelander names for particularly noteworthy
sources.

Next, on-going industrial synoptic surveys will undoubtedly discover
subclasses and new classes of interesting transients.  I specifically
suggest that the IAU in cooperation with GCVS  hold a contest every
year and dole out the remaining classic Argelander designations to
to exotic and new subclasses of transients.  The contest should be
advertised widely and all on-going time domain surveys should be
encouraged to send in nominations for spectacular variables uncovered
by their surveys.  It is well known that prizes catalyze activity
(cf.\ the $X$PRIZE). Thus the contests will certainly catalyze
research in variable star astronomy and likely have a halo effect
and bring attention of the larger astronomical community to stellar
astronomy. Perhaps a patron of astronomy could be persuaded to
underwrite the annual gala prize ceremony to be held either at
Moscow or Bonn!

Observers using Northern facilities may worry that most of the
Argelander slots in the Northern constellations have been used up.
Fortunately, as can be seen from Figure~\ref{fig:CumulativeNA} there
still remain significant number of slots that Northern facilities
can strive for.


\acknowledgements This report grew out of a during-the-dinner
conversation at a recently concluded PTF-Theory Network meeting. I
am grateful to Edwin Henneken, IT specialist, ADS, Center for
Astrophysics, Harvard University for helping me understand the
language  of machine queries to ADS; Sterl Phinney for excellent
comments; Howard Bond for catching a number of embarrassingly
elementary errors and typographical mistakes; Lynne Hillenbrand for
bringing to my attention the important paper by Townsely;  Trevor
David for his contribution to \S\ref{sec:Argelander}; Anna Ho for
a most careful reading; and N.\ Samus for educating me about GCVS
and VSX.  I would like to thank Bruce Margon  and Virginia Trimble
for their enthusiasm for this sort of work and Chris Bochanek for
feedback of an earlier version. Finally, this informal article would
not have been possible without the selfless work of librarians,
software engineers and astronomers at ADS and SIMBAD.

\IfFileExists{./TopArgelander.bbl}
{\input{TopArgelander.bbl}}
{\bibliographystyle{aasjournal}
\bibliography{bibTopArgelander}}

\appendix

\section{Naming Schemes}
\label{sec:NamingSchemes}

The first experience I had with naming scheme was on the eve of the
commissioning of  the Palomar Transient Factory \citep{lkd+09}.
This was a project aimed at a systematic study of the variable and
transient optical sky. The eponymous project was expected to churn
out a large number of transients.  The question arose on the naming
scheme for PTF discovered transients. The ensuing discussion was
intense and emotional. In the end the collaboration voted to continue
the existing scheme for supernovae in which the A--Z, AA--AZ,
BB--BZ,..,AAA-AAZ serve as postfixes to the year of discovery (e.g.\
SN\,1987A) except we dropped the number of centuries, e.g.\ PTF\,09uj and
iPTF\,14atg.

My next experience arose in the field of gamma-ray bursts (GRBs).
The traditional naming scheme is GRB\,{\it yymmdd} where {\it yy}
is the last two digits of the year, {\it mm}  is the month number
(January=1, December=12) and {\it dd} is the UT day number (1, 2,
.. 31).  Astronomers working in this field immediately recognize,
for example, GRB\,970228, GRB\,970508, GRB\,98024, GRB\,990123 and
GRB\,030329. This scheme served well for quite some time until the
rate of detection of GRBs increased to the point where more than
one GRB was discovered in a single UT day. At this point, there
were two options: use fractional day to distinguish one GRB from
another that happened on the same day or adopt the supernova
convention. I voted for the former (and felt strongly about it).
In the end the latter convention was adopted.

In retrospect I think the decision that was adopted was the better
choice. To start with the GRB rate is probably no more than 1,000
per year (REF) or $\lambda= 2.73$\,day$^{-1}$.  Thus the chance of
detecting $>[2,4,6,8,10]$ GRBs per day by an {\it all-sky} detector
is $[0.76, 0.29, 0.06, 0.007,0.0056]$, respectively. For a detector
which covers say a fourth of the sky but at sufficient sensitivity
to probe the faint end (a more reasonable prospect) the probabilities
for $>[2,3,4,5]$ are $[0.15, 0.03, 0.005, 0.0007]$, respectively.
Thus the adopted scheme has sufficient granularity to accommodate
days when the celestial sky is particularly fecund with GRBs. Next,
the designation A, B, C is easier to remember, compared to, say,
GRB\,190304.15.

As noted and expanded below short names are preferred over long
names. However, if the sample size is very large then it makes
little sense to use short names. For this reason catalogs with large
number of entries (e.g.\ ROSAT catalog, the Sloan Digital Sky Survey)
have adopted a logically constructed name (as in RX\,J0806.3+1527).

My above  experiences made me investigate naming schemes in other
areas of human activity.  I quote two examples for which considerable
thought was given to the naming schemes, though with entirely
different goals in mind.

\subsubsection*{A1. Hurricanes.}

In some regions of the world hurricanes have a major impact.
Consequently the National Hurricane center has given considerable
thought on how hurricanes are
named\footnote{\url{http://www.nhc.noaa.gov/aboutnames.shtml}}:
``Experience shows that the use of short, distinctive given names
in written as well as spoken communications is quicker and less
subject to error than the older more cumbersome latitude-longitude
identification methods."

The Center maintains six lists of 21 names (which featured only
names of women until 1979) and these are recycled every six years.
When the number of hurricanes exceeds 21 the greek alphabets are
invoked ($\alpha$, $\beta$, ...).   Hopefully, global warming will
not result in the usage of Sanskrit, Hebrew and Cyrillic alphabets.
Fortunately, given the correlation length of hurricanes  it is
unlikely that we will have to face this particular concern.

\subsubsection*{A2. Moody's Rating Scheme}

In the financial world, Moody's is a well known financial
rating company.\footnotemark\footnotetext{See
\texttt{http://www.moodys.com}.} The adjective ``well known'' is
simply a statement about the perceived standing of Moody's.  The
fortunes of companies and even countries are tied to the Moody's
rating.  If you come down a notch in the Moody's rating then you
could be losing say a few billion to a few trillion dollars of
perceived wealth. In fact, some of my colleagues who planned to
retire around 2008 had to rapidly revise their future plans owing
to the considerable losses in the market, aided in part by 
(false) high ratings given to funds based on mortgage funds.

Given the burden carried by Moody's one would expect great financial
and mathematical sophistication on their part. However, one glance
at their rating terminology should keep you wondering whether the
next global financial meltdown is round the corner.

\begin{enumerate}
\item{} {\bf A series: Aaa, Aa1, Aa2, Aa3, A1, A2, A3.}
These ratings include the preferred stocks and bonds ranging
from ``gilt edged'' to favorable investments.

\item{}{\bf Baa1, Baa2, Baa3, Ba1, Ba2, Ba3 B1, B2, B3.}
These range from adequate investments to investments with some
risks.

\item{}{\bf Caa1, Cass2, Caa3, Ca, C.}
Poor prospects.

\end{enumerate}

It is one thing for astronomers to revel in DQ\,Hercules and
PTF\,11kly but an entirely different thing for the world to trust
Moody's ratings which appear to be based on an  arcane and obscure
naming scheme.

\section{VV Andromedae}
\label{sec:VVAnd}

Upon my drawing attention to the  curious problem of VV\,And, both
Sterl Phinney, California Institute of Technology and Howard Bond,
Space Telescope Science Institute, investigated and independently
came to the same conclusion. Here is paraphrased  report from
Phinney: ``I believe I have solved the mystery of VV Andromedae. A
Google search shows that VV Andromedae was reported in {\it Popular
Astronomy}.\footnote{Popular Astronomy, Volume 21 (1913).  Linked
to "recently discovered variable stars" from Astronomische Nachrichten
\#4669.} VV And is star 110 in the list of newly assigned designations,
giving position as 23h 33m 45s +34 59 (max mag 9.7, min mag 10.2).

However a later paper \citep{pavel28} explains why it doesn't have
further data: In this note, to the limits of my German,  Pavel ({\it
ibid})  states that someone named Pra{\^c}ka claimed to have
discovered an Algol star with a period of 0.959 days, varying between
mag 9.7 and 10.2, near ST Andromedae.  However Pavel, in 29 exposures
with the 40-cm astrograph, was unable to find any variable star
down to 13--14 magnitude at or near that position other than ST
Andromedae, and concludes that if Pra\v cka observed a variable
star at all, it was certainly not in the vicinity of ST Andromedae.

So it seems VV Andromedae was given a name based on Pracka's data,
but was later determined by Pavel to have been some kind of a mistake
on Pra{\v c}ka's part.   It is not  clear why SIMBAD did not find
either of the above (under -either- VV And or ST And!), while Google
did."

I consulted Dr.\ Nikolai N.\ Samus who is the head of the group of
General Catalogue of Variable Stars, Sternberg Astronomical Institute
of Russia. He remarked ``The problem of VV And is not unique. There
were cases of repeated discoveries of the same stars, giving
variable-star names to asteroids, even to images on photographic
plates exposed twice." In fact, GCVS maintains a list of such errant
entries.

\newpage
\LongTables
\begin{deluxetable}{lllll}
\tabletypesize{\scriptsize}
\tablecaption{Top Three Argelander Stars by Constellation}
\tablewidth{0pt}
\tablehead{
\colhead{Star} & \colhead{Star} & \colhead{Star} & \colhead{Papers} & NA \\
}
\input{TableConstellations}
  \tablenotetext{}{Columns (1-3): Star names. Columns (4): The
  corresponding numbers of papers which refer to the stars. Column
  (5): number of classic Argelander designations that have not yet
  been assigned to variable stars. A ``-" indicates that all
  Argelander designations have been used up.}
 \label{tab:TopThree}
\end{deluxetable}

\end{document}

%% file: TableTopStars.tex
\startdata
CM Tau & 4346 & 188271\\
CW Leo & 1941 & 88253\\
HU Vel & 1815 & 77437\\
HZ Her & 1801 & 76800\\
BL Lac & 1772 & 169329\\
BW Tau & 1663 & 112243\\
GP Vel & 1247 & 53917\\
T Tau & 1183 & 91083\\
SS Cyg & 1141 & 40155\\
TW Hya & 1063 & 55204\\
U Gem & 986 & 42611\\
AD Leo & 973 & 39306\\
ZZ Lep & 956 & 40928\\
DQ Her & 907 & 35728\\
VY CMa & 892 & 48668\\
AM Her & 879 & 40961\\
HL Tau & 851 & 55187\\
WZ Sge & 845 & 32091\\
X Per & 835 & 42657\\
R Leo & 829 & 32662\\
RS Oph & 824 & 22712\\
AB Dor & 823 & 43248\\
RR Lyr & 821 & 90775\\
DG Tau & 819 & 54421\\
GK Per & 802 & 27134\\
AB Aur & 781 & 44379\\
YY Gem & 764 & 42479\\
AU Mic & 739 & 32788\\
UV Cet & 737 & 33829\\
R CrB & 733 & 23525\\
QX Nor & 713 & 34919\\
CH Cyg & 710 & 16842\\
EZ CMa & 707 & 34578\\
EV Lac & 701 & 28666\\
IL Aqr & 697 & 47955\\
YZ CMi & 689 & 27718\\
S Mon & 688 & 52156\\
W Com & 684 & 44964\\
R Aqr & 683 & 25963\\
CF UMa & 679 & 56941\\
AR Lac & 677 & 29520\\
AU CVn & 667 & 48741\\
BR Cir & 662 & 28901\\
RS CVn & 657 & 50874\\
II Peg & 639 & 26592\\
GU Mus & 637 & 32034\\
R Cas & 636 & 23675\\
UX Ari & 633 & 31696\\
RW Aur & 621 & 35480\\
FU Ori & 618 & 34290
\enddata

%% file: TableTopCites.tex
\startdata
CM Tau & 188271 & 4346\\
BL Lac & 169329 & 1772\\
BW Tau & 112243 & 1663\\
T Tau & 91083 & 1183\\
RR Lyr & 90775 & 821\\
CW Leo & 88253 & 1941\\
HU Vel & 77437 & 1815\\
HZ Her & 76800 & 1801\\
CF UMa & 56941 & 679\\
TW Hya & 55204 & 1063\\
HL Tau & 55187 & 851\\
DG Tau & 54421 & 819\\
GP Vel & 53917 & 1247\\
S Mon & 52156 & 688\\
RS CVn & 50874 & 657\\
AU CVn & 48741 & 667\\
VY CMa & 48668 & 892\\
IL Aqr & 47955 & 697\\
GG Tau & 46093 & 540\\
W Com & 44964 & 684\\
AB Aur & 44379 & 781\\
AB Dor & 43248 & 823\\
GM Aur & 42856 & 459\\
X Per & 42657 & 835\\
U Gem & 42611 & 986\\
YY Gem & 42479 & 764\\
AP Lib & 42050 & 494\\
AM Her & 40961 & 879\\
BP Tau & 40938 & 536\\
ZZ Lep & 40928 & 956\\
SS Cyg & 40155 & 1141\\
AE Aur & 39624 & 466\\
CN Leo & 39466 & 551\\
RY Tau & 39422 & 580\\
AD Leo & 39306 & 973\\
AA Tau & 37966 & 489\\
DQ Her & 35728 & 907\\
RW Aur & 35480 & 621\\
SU Aur & 35346 & 486\\
QX Nor & 34919 & 713\\
EZ CMa & 34578 & 707\\
FU Ori & 34290 & 618\\
UV Cet & 33829 & 737\\
AU Mic & 32788 & 739\\
R Leo & 32662 & 829\\
WZ Sge & 32091 & 845\\
GU Mus & 32034 & 637\\
UX Ari & 31696 & 633\\
W UMa & 30290 & 602\\
AR Lac & 29520 & 677
\enddata

%% file: TableConstellations.tex
\startdata
Z And & GX And & RT And &  478,  377, 347 & 1 \\
AG Ant & BW Ant & U Ant &  263,  141, 123 & 227 \\
MY Aps & S Aps & NN Aps &  109,  107, 79 & - \\
IL Aqr & R Aqr & AE Aqr &  697,  683, 554 & 13 \\
R Aql & FF Aql & RR Aql &  518,  317, 253 & - \\
S Ara & AE Ara & R Ara &  84,  65, 63 & - \\
UX Ari & TT Ari & X Ari &  633,  398, 236 & 219 \\
AB Aur & RW Aur & SU Aur &  781,  621, 486 & - \\
RX Boo & HN Boo & HP Boo &  389,  218, 217 & - \\
RR Cae & R Cae & X Cae &  106,  79, 65 & 309 \\
Z Cam & AX Cam & SV Cam &  452,  353, 337 & - \\
RS Cnc & R Cnc & X Cnc &  279,  266, 244 & 64 \\
AU CVn & RS CVn & Y CVn &  667,  657, 441 & 137 \\
VY CMa & EZ CMa & Z CMa &  892,  707, 480 & - \\
YZ CMi & CY CMi & BG CMi &  689,  284, 189 & 172 \\
BY Cap & BB Cap & RT Cap &  158,  128, 103 & 212 \\
AG Car & OY Car & HR Car &  552,  525, 271 & - \\
R Cas & RZ Cas & SU Cas &  636,  416, 389 & - \\
V Cen & BV Cen & XX Cen &  217,  162, 152 & - \\
VW Cep & U Cep & VV Cep &  499,  447, 411 & - \\
UV Cet & BE Cet & FS Cet &  737,  404, 331 & 112 \\
Z Cha & CU Cha & DX Cha &  564,  355, 276 & 97 \\
BR Cir & BW Cir & AX Cir &  662,  151, 108 & 166 \\
TV Col & TX Col & T Col &  278,  126, 85 & 251 \\
W Com & LS Com & FK Com &  684,  392, 327 & 18 \\
R CrA & TY CrA & S CrA &  409,  249, 237 & - \\
R CrB & T CrB & TZ CrB &  733,  589, 450 & 230 \\
TY Crv & R Crv & W Crv &  110,  76, 69 & 281 \\
TV Crt & R Crt & SV Crt &  297,  180, 175 & 265 \\
BP Cru & BZ Cru & S Cru &  545,  235, 157 & 158 \\
SS Cyg & CH Cyg & X Cyg &  1141,  710, 414 & 1 \\
HR Del & NT Del & EU Del &  437,  318, 151 & 17 \\
AB Dor & S Dor & R Dor &  823,  307, 213 & 249 \\
BY Dra & AG Dra & CM Dra &  565,  451, 360 & - \\
S Equ & SY Equ & U Equ &  116,  78, 54 & 298 \\
EP Eri & EF Eri & DO Eri &  418,  394, 345 & 51 \\
UZ For & R For & TZ For &  223,  164, 89 & 255 \\
U Gem & YY Gem & OU Gem &  986,  764, 207 & 1 \\
BP Gru & RS Gru & S Gru &  97,  85, 66 & 183 \\
HZ Her & DQ Her & AM Her &  1801,  907, 879 & - \\
R Hor & TW Hor & WW Hor &  156,  116, 70 & 270 \\
TW Hya & EX Hya & W Hya &  1063,  606, 524 & - \\
VW Hyi & WX Hyi & BL Hyi &  583,  186, 184 & 198 \\
CI Ind & T Ind & CD Ind &  112,  75, 70 & 207 \\
BL Lac & EV Lac & AR Lac &  1772,  701, 677 & - \\
CW Leo & AD Leo & R Leo &  1941,  973, 829 & 68 \\
RW LMi & SV LMi & R LMi &  343,  309, 297 & 270 \\
ZZ Lep & R Lep & SS Lep &  956,  285, 188 & 248 \\
AP Lib & HO Lib & KX Lib &  494,  424, 403 & - \\
IL Lup & RU Lup & EX Lup &  360,  316, 206 & - \\
EI Lyn & RR Lyn & AE Lyn &  181,  178, 157 & 147 \\
RR Lyr & R Lyr & MV Lyr &  821,  287, 225 & - \\
TU Men & YY Men & TZ Men &  140,  115, 107 & 251 \\
AU Mic & AT Mic & AX Mic &  739,  296, 179 & 203 \\
S Mon & R Mon & T Mon &  688,  474, 414 & - \\
GU Mus & KR Mus & KN Mus &  637,  463, 275 & - \\
QX Nor & QV Nor & S Nor &  713,  338, 312 & - \\
CL Oct & DR Oct & UV Oct &  121,  95, 82 & 162 \\
RS Oph & U Oph & Y Oph &  824,  343, 325 & 1 \\
FU Ori & U Ori & BM Ori &  618,  456, 306 & - \\
AR Pav & S Pav & Y Pav &  137,  73, 70 & - \\
II Peg & AG Peg & EQ Peg &  639,  466, 390 & - \\
X Per & GK Per & MX Per &  835,  802, 345 & - \\
SX Phe & AE Phe & AI Phe &  291,  103, 97 & 203 \\
RR Pic & VZ Pic & AK Pic &  289,  230, 108 & 260 \\
TX Psc & ZZ Psc & WX Psc &  415,  384, 370 & 110 \\
TW PsA & HU PsA & TY PsA &  234,  104, 90 & 275 \\
QX Pup & VV Pup & RS Pup &  463,  373, 284 & - \\
T Pyx & TY Pyx & VW Pyx &  334,  239, 104 & 192 \\
R Ret & S Ret & TT Ret &  58,  58, 45 & 286 \\
WZ Sge & QX Sge & QV Sge &  845,  605, 516 & - \\
VX Sgr & U Sgr & RY Sgr &  510,  404, 346 & 1 \\
U Sco & AK Sco & RV Sco &  413,  183, 161 & - \\
R Scl & BB Scl & VY Scl &  253,  130, 128 & 211 \\
R Sct & RY Sct & EV Sct &  276,  259, 210 & - \\
NP Ser & MQ Ser & MM Ser &  558,  291, 275 & - \\
AY Sex & SW Sex & RW Sex &  170,  168, 163 & 237 \\
CM Tau & BW Tau & T Tau &  4346,  1663, 1183 & - \\
RR Tel & PZ Tel & QS Tel &  496,  199, 108 & - \\
RW Tri & X Tri & R Tri &  272,  205, 175 & 212 \\
KZ TrA & MM TrA & R TrA &  480,  196, 152 & - \\
CF Tuc & W Tuc & BS Tuc &  205,  88, 79 & 174 \\
CF UMa & KV UMa & W UMa &  679,  615, 602 & - \\
RR UMi & S UMi & U UMi &  154,  117, 108 & 276 \\
HU Vel & GP Vel & IM Vel &  1815,  1247, 238 & - \\
GW Vir & CU Vir & EQ Vir &  437,  310, 296 & - \\
UY Vol & R Vol & AI Vol &  498,  70, 36 & 268 \\
QZ Vul & SV Vul & ER Vul &  380,  372, 323 & - 
\enddata

%% file: TopArgelander.bbl
\begin{thebibliography}{}
\expandafter\ifx\csname natexlab\endcsname\relax\def\natexlab#1{#1}\fi


\bibitem[{{Abbott} {et~al.}(2016){Abbott}, {Abbott}, {Abbott}, {Abernathy},
  {Acernese}, \& et~al.}]{2016ApJ...818L..22A}
{Abbott}, B.~P., {Abbott}, R., {Abbott}, T.~D., {et~al.} 2016, \apjl, 818, L22

\bibitem[{{Argelander} \& {Cannon}(1912)}]{1912PA.....20...91A}
{Argelander}, F.~W.~A., \& {Cannon}, A.~J. 1912, Popular Astronomy, 20, 91

\bibitem[{{Barros} {et~al.}(2007){Barros}, {Marsh}, {Dhillon}, {Groot},
  {Littlefair}, {Nelemans}, {Roelofs}, {Steeghs}, \& {Wheatley}}]{bmd+07}
{Barros}, S.~C.~C., {Marsh}, T.~R., {Dhillon}, V.~S., {et~al.} 2007, \mnras,
  374, 1334

\bibitem[{{Burwitz} \& {Reinsch}(1999)}]{br99}
{Burwitz}, V., \& {Reinsch}, K. 1999, in Astronomische Gesellschaft Abstract
  Series, Vol.~15, Astronomische Gesellschaft Abstract Series, ed. R.~E.
  {Schielicke}

\bibitem[{{Israel} {et~al.}(1999){Israel}, {Panzera}, {Campana}, {Lazzati},
  {Covino}, {Tagliaferri}, \& {Stella}}]{ipc+99}
{Israel}, G.~L., {Panzera}, M.~R., {Campana}, S., {et~al.} 1999, \aap, 349, L1

\bibitem[{{Law} {et~al.}(2009){Law}, {Kulkarni}, {Dekany}, {Ofek}, {Quimby},
  {Nugent}, {Surace}, {Grillmair}, {Bloom}, {Kasliwal}, {Bildsten}, {Brown},
  {Cenko}, {Ciardi}, {Croner}, {Djorgovski}, {van Eyken}, {Filippenko}, {Fox},
  {Gal-Yam}, {Hale}, {Hamam}, {Helou}, {Henning}, {Howell}, {Jacobsen},
  {Laher}, {Mattingly}, {McKenna}, {Pickles}, {Poznanski}, {Rahmer}, {Rau},
  {Rosing}, {Shara}, {Smith}, {Starr}, {Sullivan}, {Velur}, {Walters}, \&
  {Zolkower}}]{lkd+09}
{Law}, N.~M., {Kulkarni}, S.~R., {Dekany}, R.~G., {et~al.} 2009, \pasp, 121,
  1395

\bibitem[{{Ma} {et~al.}(2015){Ma}, {Kasen}, {Hopkins}, {Faucher-Gigu{\`e}re},
  {Quataert}, {Kere{\v s}}, \& {Murray}}]{mkh+15}
{Ma}, X., {Kasen}, D., {Hopkins}, P.~F., {et~al.} 2015, \mnras, 453, 960

\bibitem[{{Pavel}(1928)}]{pavel28}
{Pavel}, F. 1928, Astronomische Nachrichten, 233, 367

\bibitem[{{Townley}(1915)}]{1915PASP...27..209T}
{Townley}, S.~D. 1915, \pasp, 27, 209

\end{thebibliography}
